\documentclass[12pt,a4paper]{article}

\pdfoutput=1

\usepackage[DIV13,BCOR0mm]{typearea}

\usepackage{natbib}
\bibpunct[, ]{(}{)}{,}{a}{}{,}

\usepackage{graphicx}

\newcommand{\beq}{\begin{equation}}
\newcommand{\eeq}{\end{equation}}
\newcommand{\beqa}{\begin{eqnarray}}
\newcommand{\eeqa}{\end{eqnarray}}
\newcommand{\beqan}{\begin{eqnarray*}}
\newcommand{\eeqan}{\end{eqnarray*}}

\newcommand{\bigfrac}[2]{\mbox{$\displaystyle\frac{#1}{#2}$}}

\newcommand{\D}{\mathrm{d}}

\newcommand{\eps}{\varepsilon}

\newcommand{\pabl}[2]{\frac{\partial #1}{\partial #2}}

\newcommand{\nl}{\nonumber\\}

\newcommand{\degN}{\ensuremath{^\circ\mathrm{N}}}    
\newcommand{\degS}{\ensuremath{^\circ\mathrm{S}}}    

\def\pmb#1{\setbox0=\hbox{#1}%
  \kern-.025em\copy0\kern-\wd0
  \kern.05em\copy0\kern-\wd0
  \kern-.025em\raise.0433em\box0}

\def\pmbm#1{\setbox0=\hbox{$#1$}%
  \kern-.025em\copy0\kern-\wd0
  \kern.05em\copy0\kern-\wd0
  \kern-.025em\raise.0433em\box0}

\def\vecbi#1{\relax\ifmmode\mathchoice
  {\mbox{\boldmath$\relax\displaystyle#1$}}
  {\mbox{\boldmath$\relax\textstyle#1$}}
  {\mbox{\boldmath$\relax\scriptstyle#1$}}
  {\mbox{\boldmath$\relax\scriptscriptstyle#1$}}\else
  \hbox{\boldmath$\relax\textstyle#1$}\fi} 

\def\vecbu#1{\relax\ifmmode\mathchoice
  {\mbox{\boldmath$\bf\displaystyle#1$}}
  {\mbox{\boldmath$\bf\textstyle#1$}}
  {\mbox{\boldmath$\bf\scriptstyle#1$}}
  {\mbox{\boldmath$\bf\scriptscriptstyle#1$}}\else
  \hbox{\boldmath$\bf\textstyle#1$}\fi}    

\def\tenssu#1{\relax\ifmmode\mathchoice
    {\mbox{$\sf\displaystyle#1$}}
    {\mbox{$\sf\textstyle#1$}}
    {\mbox{$\sf\scriptstyle#1$}}
    {\mbox{$\sf\scriptscriptstyle#1$}}\else
    \hbox{$\sf\textstyle#1$}\fi}           

\begin{document}

\title{\textbf{MAIC-2, a latitudinal model for the Martian surface temperature,
               atmospheric water transport and surface glaciation}}

\author{\textsc{Ralf Greve}\thanks{E-mail: greve@lowtem.hokudai.ac.jp}\\[0.5ex]
        {\normalsize Institute of Low Temperature Science, Hokkaido University,}\\[-0.25ex]
        {\normalsize Kita-19, Nishi-8, Kita-ku, Sapporo 060-0819, Japan}\\[1.5ex]
        \textsc{Bj\"orn Grieger}\\[0.5ex]
        {\normalsize European Space Astronomy Centre, P.O.\ Box -- Apdo.\ de Correos 78,}\\[-0.25ex]
        {\normalsize 28691 Villanueva de la Ca\~nada, Madrid, Spain}\\[1.5ex]
        \textsc{Oliver J.~Stenzel}\\[0.5ex]
        {\normalsize Max Planck Institute for Solar System Research,}\\[-0.25ex]
        {\normalsize Max-Planck-Stra\ss{}e 2, 37191 Katlenburg-Lindau, Germany}}

\date{}

\maketitle

\begin{abstract}

\vspace*{-149mm}

\noindent\hspace*{-10mm}{\footnotesize\emph{Planetary and Space Science}
         \textbf{58}~(6), 931--940 (2010)
\\[-0.3ex]\hspace*{-10mm}doi: 10.1016/j.pss.2010.03.002
\\{}\hspace*{-10mm}--- Authors' version ---}

\vspace*{136mm}

The Mars Atmosphere-Ice Coupler MAIC-2 is a simple, latitudinal
model, which consists of a set of parameterisations for the
surface temperature, the atmospheric water transport and the
surface mass balance (condensation minus evaporation) of water
ice. It is driven directly by the orbital parameters obliquity,
eccentricity and solar longitude ($L_\mathrm{s}$) of
perihelion. Surface temperature is described by the Local
Insolation Temperature (LIT) scheme, which uses a daily and
latitude-dependent radiation balance. The evaporation rate of
water is calculated by an expression for free convection,
driven by density differences between water vapor and ambient
air, the condensation rate follows from the assumption that any
water vapour which exceeds the local saturation pressure
condenses instantly, and atmospheric transport of water vapour
is approximated by instantaneous mixing. Glacial flow of ice
deposits is neglected. Simulations with constant orbital
parameters show that low obliquities favour deposition of ice
in high latitudes and vice versa. A transient scenario driven
by a computed history of orbital parameters over the last
10~million years produces essentially monotonically growing
polar ice deposits during the most recent 4~million years, and
a very good agreement with the observed present-day polar
layered deposits. The thick polar deposits sometimes continue
in thin ice deposits which extend far into the mid latitudes,
which confirms the idea of ``ice ages'' at high obliquity.

\end{abstract}

\section{Introduction}
\label{sect_intro}

\begin{figure}[htb]
  \centering
  \includegraphics[width=130mm]{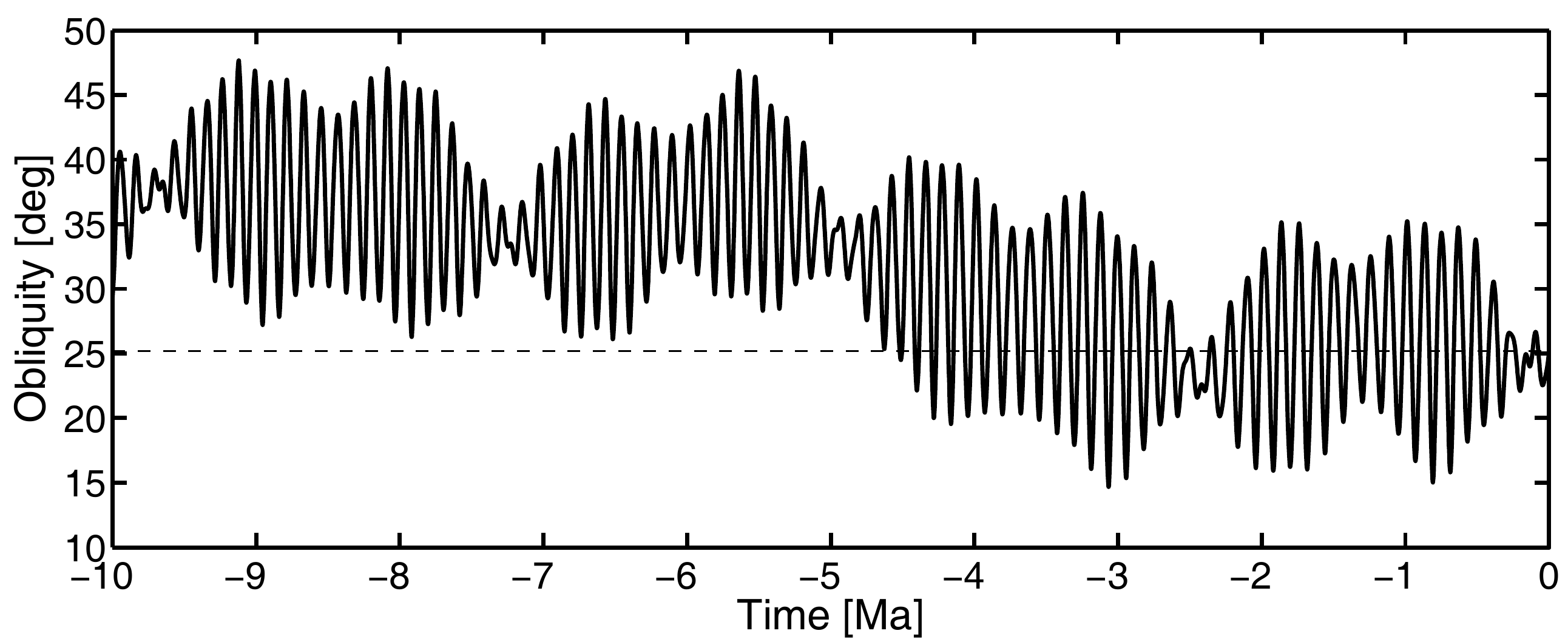}
  \par\vspace*{-1ex}\par
  \caption{Martian obliquity for the last 10~Ma
           \citep{laskar_etal_04}.}
  \label{fig_obliq}
\end{figure}

On time scales of $10^5$--$10^6$ years, Mars has experienced
large periodic changes of the orbital elements obliquity,
eccentricity and equinox precession. These changes have an
impact on the Martian climate. The obliquity determines the
strength of the seasons and the latitudinal distribution of
mean solar insolation. The eccentricity determines the
magnitude of the asymmetry of insolation with season, and the
equinox precession determines the timing of the asymmetry of
solar insolation with season. On Earth, the so-called
Milankovitch cycles of much weaker orbital changes with periods
of 20, 40 and 100~ka are considered driving forces for climate
variations like the glacial/interglacial cycles. It can,
therefore, be expected that the main Martian $\pm{}10^\circ$
obliquity cycles with periods of 125~ka and 1.3~Ma and the
secular shift from high to low average obliquities at 4--5~Ma
ago \citep[][shown in Fig.~\ref{fig_obliq}]{laskar_etal_04}
have significant impacts on the climate and the polar layered
deposits (PLDs) due to large insolation changes in the polar
regions.

This idea is supported by the presence of light-dark layers in
the PLDs, which are exposed in the surface troughs and close to
the margins, and which are actually the reason for the term
``polar \emph{layered} deposits''. These layers indicate a
strongly varying dust content of the ice due to varying
climatic conditions in the past. During periods of high
obliquities, insolation in the polar regions is large, which
entails higher sublimation rates of superficial ice of the PLDs
and probably of permafrost in the ground. This may lead to the
formation of a thicker and dustier atmosphere, so that dust
accumulates on the PLDs. By contrast, during periods of low
obliquities, the atmosphere is thin and dust deposition is low,
so that clean ice forms at the surface of the PLDs. Along this
line of reasoning, \citet{head_etal_03} presented evidence for
past glaciation in the mid-latitudes and suggested that Mars
experienced ``ice ages'' during periods of high obliquity like
that from about 2.1 to 0.4~Ma ago (with obliquity maxima of
${}\approx{}35^\circ$). These ice ages were supposedly
characterised by warmer polar climates, enhanced mass loss of
the PLDs due to sublimation and the formation of metres-thick
ice deposits equatorward to approximately $30\degN$/S.

In a number of studies, General Circulation Models (GCMs) have
been applied to the Martian atmosphere
\citep[e.g.,][]{pollack_etal_90, read_etal_97, forget_etal_99,
richardson_wilson_02, haberle_etal_03, takahashi_etal_03}.
These models, all derivatives of Earth GCMs, solve the
equations of fluid dynamics and thermodynamics and include
e.g.\ the processes of radiative transfer, cloud formation,
regolith-atmosphere water exchange, and advective transport of
dust and trace gases. However, they have essentially been
designed to simulate the present-day atmosphere in as much
detail as possible, and thus are computationally too expensive
to permit long-term paleoclimate studies.
\citet{segschneider_etal_05} and \citet{stenzel_etal_07}
adapted an Earth System Model of Intermediate Complexity (EMIC)
to Mars. This ``Planet Simulator Mars'' (PlaSim-Mars, formerly
called ``Mars Climate Simulator'') allows in principle
simulations over longer, climatological time scales. So far,
only scenarios for present-day conditions and varied obliquity
angles have been considered, and the impact on the PLDs has
been studied by coupling PlaSim-Mars with the
three-dimensional, dynamic/thermodynamic ice-sheet model
SICOPOLIS (http://sicopolis.greveweb.net/). In addition to
that, simple one-di\-men\-sio\-nal models have been used to
study specific processes that do not require full solution of
the dynamic equations. Examples are the radiative transfer
model by \citet{gierasch_goody_68}, the energy balance model by
\citet{armstrong_etal_04}, regolith-atmosphere water exchange
\citep{jakosky_85}, and formation of water ice clouds
\citep{michelangeli_etal_93}.

In this study we aim at simulating the surface glaciation of
the entire planet with a simple model that depends only on
latitude and time. This model, termed the Mars Atmosphere-Ice
Coupler Version 2, or MAIC-2 in short, consists of a set of
parameterisations for the surface temperature, the atmospheric
water transport and the surface mass balance (condensation
minus evaporation) of water ice. It is driven directly by the
orbital parameters obliquity, eccentricity and solar longitude
($L_\mathrm{s}$) of perihelion, which were published by
\citet{laskar_etal_04} for the period from 20 million years ago
until 10 million years into the future. MAIC-2 is applied to
two different kinds of scenarios, namely (i) scenarios with
orbital parameters kept constant over time, and (ii) transient
scenarios forced by the history of orbital parameters over the
last 10 million years. The evolution of surface glaciation is
studied for these scenarios under the simplifying assumption of
negligible glacial flow, so that changes of local ice thickness
are exclusively determined by the local surface mass balance
provided by MAIC-2.

\section{Model MAIC-2}
\label{sect_maic2}

\begin{figure}[htb]
  \centering
  \includegraphics[width=80mm]{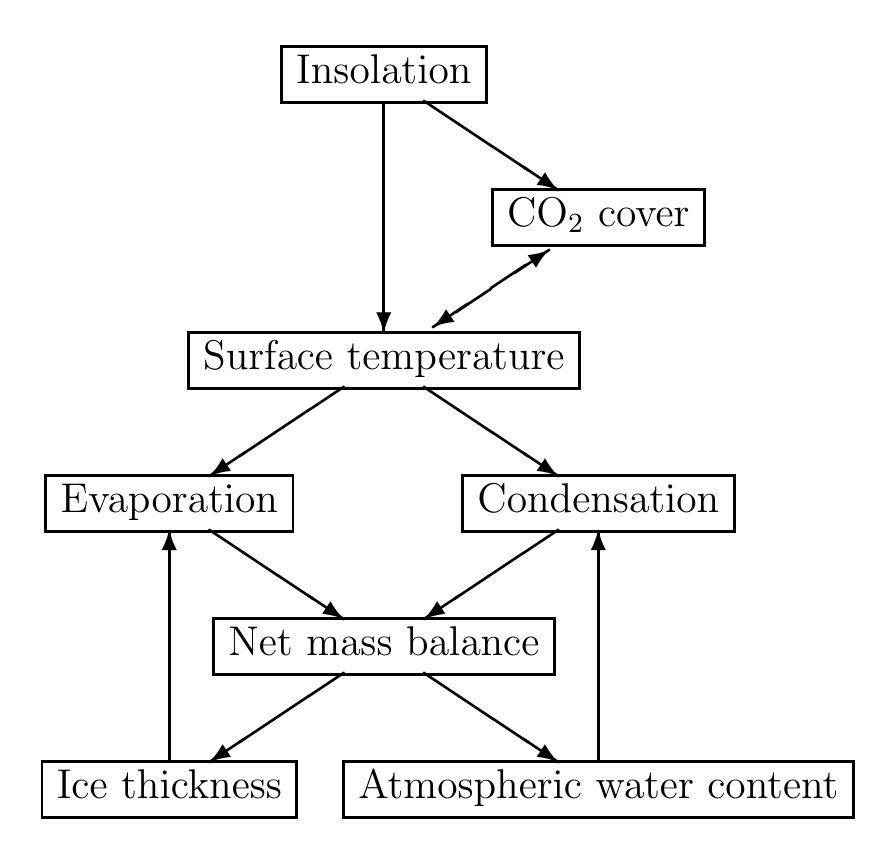}
  \par\vspace*{-1ex}\par
  \caption{Schematic illustration of the quantities and processes
  modelled by MAIC-2.}
  \label{fig_maic2_schematics}
\end{figure}

The design of MAIC-2 is illustrated schematically in
Fig.~\ref{fig_maic2_schematics}. All quantities are latitude
($\varphi$) and time ($t$) dependent, with the exception of the
atmospheric water content. Since instantaneous mixing is
assumed, only the (time dependent) global mean water content is
modelled. The formulation of the different processes is
detailed in the following sections.

\subsection{Timekeeping and orbital position}
\label{ssect_time}

We use the timekeeping of PlaSim-Mars (see
Sect.~\ref{sect_intro}), in which a Martian year consists of 12
months of 56 days (sols) length. A day has 24 hours of 61.5
minutes length, and a minute is 60 SI seconds long.

In order to compute the relation between solar longitude
$L_\mathrm{s}$ and time $t$, let us identify the beginning of a
Martian year with the northern hemisphere vernal equinox
($L_\mathrm{s}=0^\circ$). The orbit of Mars around the sun is
described by the ellipse
\beq
  r(\psi) = \frac{p}{1+\eps\cos\psi}\,,
  \label{eq_orbit}
\eeq
where $r$ is the distance Sun -- Mars, $\psi$ the orbital
position measured from the perihelion (``true anomaly''), $p$
the semi-latus rectum and $\eps$ the eccentricity. By combining
this equation and the conservation of angular momentum $l$,
\beq
  l = m r^2 \dot{\psi}
\eeq
(where $m$ denotes the mass of Mars), we find
\beq
  \dot{\psi}
  = \frac{l}{mr^2}
  = \frac{l}{mp^2}\,(1+\eps\cos\psi)^2
  = \Omega\,(1+\eps\cos\psi)^2
  \qquad
  \Big( \Omega := \frac{l}{mp^2} \Big)\,.
  \label{eq_anomaly}
\eeq
Let $L_\mathrm{s,p}$ be the solar longitude of perihelion, then
\beq
  \psi = L_\mathrm{s} - L_\mathrm{s,p}\,,
\eeq
and since $L_\mathrm{s,p}$ varies only slowly over time, we can
rewrite Eq.~(\ref{eq_anomaly}) as
\beq
  \dot{L}_\mathrm{s}
  = \Omega\,[1+\eps\cos(L_\mathrm{s}-L_\mathrm{s,p})]^2\,.
  \label{eq_ls}
\eeq
This equation can be integrated numerically over a full Martian
year (from vernal equinox to vernal equinox) for any values of
$\eps$ and $L_\mathrm{s,p}$ by a simple Euler-forward scheme.
The initial condition is $L_\mathrm{s}=0^\circ$, and the
parameter $\Omega$ is adjusted iteratively such that after one
Martian year the orbit is closed ($L_\mathrm{s}=360^\circ$),
starting from the initial guess
$\Omega_\mathrm{init}=2\pi/\mathrm{(1\;Martian\;year)}$ [which
is the correct value for a circular orbit with $\eps=0$].

\begin{figure}[htb]
  \centering
  \includegraphics[width=100mm]{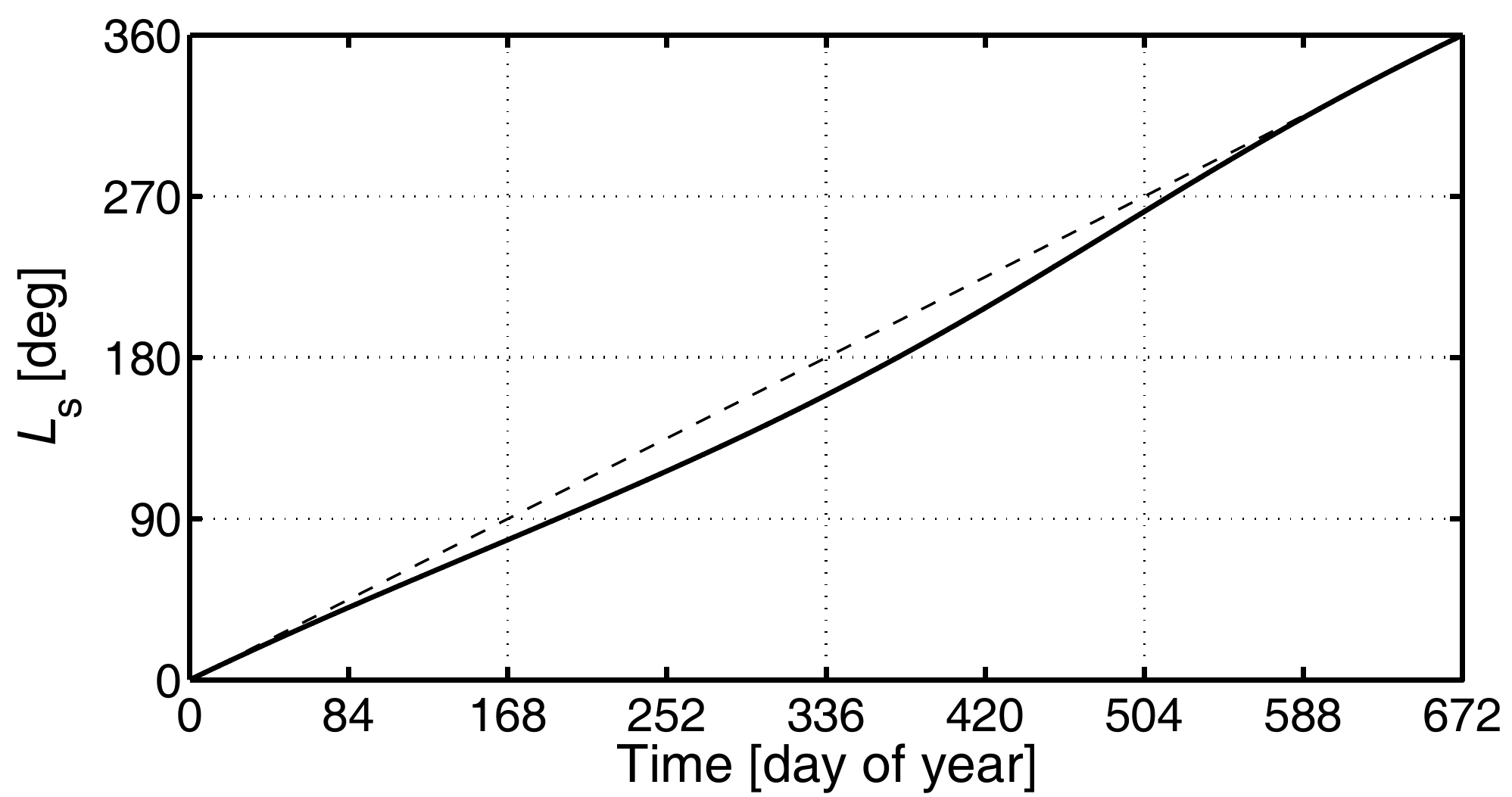}
  \par\vspace*{-1ex}\par
  \caption{Relation between solar longitude $L_\mathrm{s}$ and time $t$
  for present-day conditions (solid line). For comparison, the dashed line
  shows the linear relation for a circular orbit.}
  \label{fig_ls_time_relation}
\end{figure}

For present-day conditions ($\eps=0.093$,
$L_\mathrm{s,p}=251.0^\circ$), the result is shown in
Fig.~\ref{fig_ls_time_relation}. Since the eccentricity is much
larger for Mars than for Earth, the relation between
$L_\mathrm{s}$ and $t$ is significantly different from a linear
one. The deviation becomes as large as $21^\circ$
($L_\mathrm{s}=158.97^\circ$ instead of $180^\circ$ for day of
year 336, leading to a lag of the northern autumnal equinox by
37.7 Martian days).

\subsection{Surface temperature}
\label{ssect_temp}

In order to derive a parameterisation for the daily mean local
surface temperature $T(\varphi,t)$ (depending on latitude
$\varphi$ and time $t$), we start with the radiation balance
\beq
  \sigma T^4 = (1-A)\,F\,,
  \label{eq_rad_balance}
\eeq
where $\sigma$ is the Stefan-Boltzmann constant
($\sigma=5.67\times{}10^{-8}\,\mathrm{W\,m^{-2}\,K^{-4}}$), $A$
is the surface albedo (globally $A=0.3$ assumed) and $F$ is the
local daily mean insolation as a function of the orbital
parameters obliquity, eccentricity and solar longitude of
perihelion \citep{laskar_etal_04}. In the absence of seasonal
CO$_2$ frost, Eq.~(\ref{eq_rad_balance}) provides reasonable
results for the surface temperature. However, the equation does
not account for the fact that at
\beq
  T = T_\mathrm{cond} = \frac{b}{a-\ln P\,[\mathrm{hPa}]}
  \label{eq_CO2_condtemp}
\eeq
(where $P$ is the surface pressure, $a=23.3494$ and
$b=3182.48\,\mathrm{K}$; \citeauthor{james_etal_92}\
\citeyear{james_etal_92}) condensation of the atmospheric
CO$_2$ (formation of the seasonal ice cap) sets in. The
seasonal variation of the surface pressure is neglected, and we
use the global annual mean value $P=700\,\mathrm{Pa}$ instead.
For this value, Eq.~(\ref{eq_CO2_condtemp}) yields a
condensation temperature of
$T_\mathrm{cond}=148.7\,\mathrm{K}$. Since the atmosphere never
freezes out completely, this value constitutes the minimum of
surface temperatures which can be realised.

In order to find out when the seasonal CO$_2$ ice cap at a
given latitude $\varphi$ is present, and therefore
$T=T_\mathrm{cond}$ holds, the seasonal cap is assumed to exist
between the onset of the polar night ($t_\mathrm{dusk}$) and an
unknown time $t$ after the end of the polar night
($t_\mathrm{dawn}$). During the polar night, condensation takes
place, and the amount of formed CO$_2$ frost corresponds to the
energy (per area unit)
\beq
  W_\mathrm{cond} = \int_{t_\mathrm{dusk}}^{t_\mathrm{dawn}}
                    \sigma T_\mathrm{cond}^4\,\mathrm{d}t\,.
  \label{eq_w_cond}
\eeq
After dawn, the solar insolation causes the CO$_2$ frost to
evaporate, which requires the energy
\beq
  W_\mathrm{evap} = \int_{t_\mathrm{dawn}}^{t}
                    \left( (1-A)\,F
                    -\sigma T_\mathrm{cond}^4\right)\,\mathrm{d}t\,.
  \label{eq_w_evap}
\eeq
The time $t$ at which the CO$_2$ frost has evaporated
completely follows from
\beq
  W_\mathrm{evap} = W_\mathrm{cond}\,.
  \label{eq_w_cond_evap}
\eeq

\begin{figure}[htb]
  \centering
  \includegraphics[width=110mm]{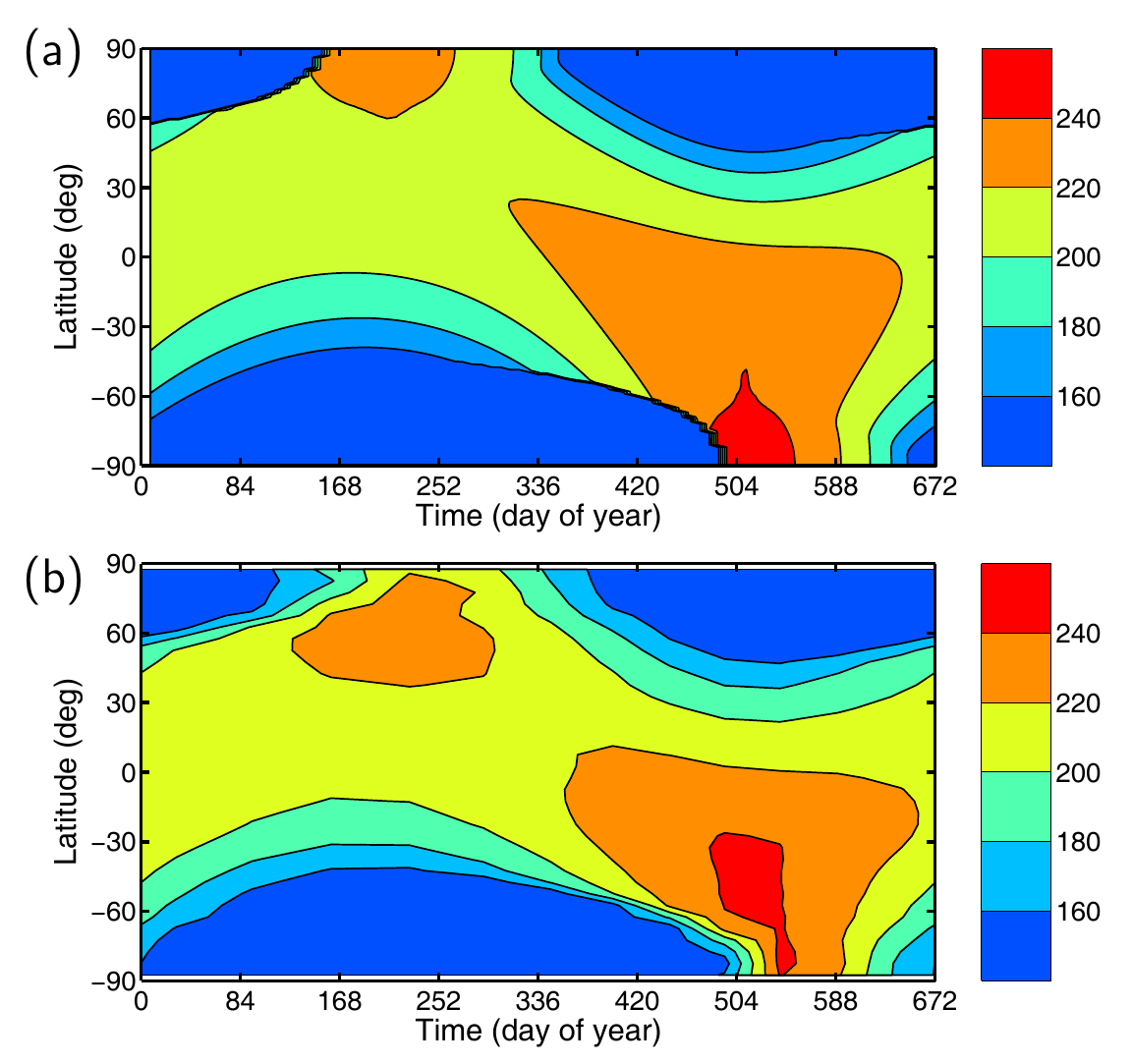}
  \par\vspace*{-1ex}\par
  \caption{(a) Daily mean surface temperature (in K) of the LIT scheme
  for present-day conditions.
  (b) Same, but from the Mars Climate Database \citep{lewis_etal_99}.}
  \label{fig_lit_temp}
\end{figure}

The scheme defined by the radiation balance
(\ref{eq_rad_balance}), modified by CO$_2$ condensation
following Eqs.~(\ref{eq_CO2_condtemp})--(\ref{eq_w_cond_evap}),
is referred to as \emph{Local Insolation Temperature scheme}
(LIT); it was first laid down by B.~Grieger (2004; talk at 2nd
MATSUP workshop, Darmstadt, Germany). The resulting daily mean
surface temperatures over one Martian year for present-day
conditions are shown in Fig.~\ref{fig_lit_temp}. They agree
well with the data given in the Mars Climate Database
\citep{lewis_etal_99}. The most notable discrepancy is that the
LIT scheme overpredicts the summer temperatures at and very
close to the poles.

The Mars Atmosphere-Ice Coupler with the LIT scheme and the
simple treatment of the surface mass balance described by
\citet{greve_etal_04a} and \citet{greve_mahajan_05} is referred
to as ``MAIC-1.5''. It was used by these authors to drive
simulations of the north polar layered deposits with the
ice-sheet model SICOPOLIS.

\subsection{Saturation pressure of water vapour}
\label{ssect_pres_sat}

The water-vapour saturation pressure $P_\mathrm{sat}$ is
obtained from the Clausius-Clapeyron relation, which can be
integrated only approximately. Different approximations are
available; we use the Magnus-Teten formula for water vapour
over ice \citep{murray_67}
\beq
  P_{\mathrm{sat}}(T) = A \exp\left( \frac{B(T-T_0)}{T-C}\right)\,,
\eeq
with $A=610.66\,\mathrm{Pa}$, $B=21.875$, $C=7.65\,\mathrm{K}$
and $T_0=273.16\,\mathrm{K}$, which has also been implemented
in the Planet Simulator Mars \citep{stenzel_etal_07}.

\subsection{Evaporation}
\label{ssect_evap}

\citet{ingersoll_70} discussed the water vapour mass flux in
the Martian carbon dioxide atmosphere. The evaporation rate $E$
of water from the surface, expressed as a mass flux in
$\mathrm{kg\,m^{-2}\,s^{-1}}$, is
\beq
  E = E_0 \times 0.17\,\Delta\eta\,\rho D\,
      \left(\frac{(\Delta\rho/\rho)\,g}{\nu^2}\right)^{1/3}\,,
  \label{eq_evap_in}
\eeq
where $E_0$ is the evaporation factor (default value equal to
unity), $\Delta\eta$ the concentration difference at the
surface and at distance, $\rho$ the atmospheric density,
$\Delta\rho$ the CO$_2$ density difference at the surface and
at distance, $D$ the diffusion coefficient of water in CO$_2$,
$g$ the acceleration due to gravity and $\nu$ the kinematic
viscosity of CO$_2$ \citep[cf.\ also][]{sears_moore_05}. The
term $\Delta\eta$ is given by
\beq
  \Delta\eta\
  = \frac{\rho_\mathrm{w}^\mathrm{sat}}{\rho}
  = \frac{M_\mathrm{w} P_\mathrm{sat}}{M_\mathrm{c} P}\,,
  \label{eq_evap_in_term1}
\eeq
where $\rho_\mathrm{w}^\mathrm{sat}$ is the saturation density
of water vapour at the surface temperature $T$ and
$M_\mathrm{w}$ and $M_\mathrm{c}$ are the molecular weights of
water and carbon dioxide, respectively. The terms $\rho$ and
$\Delta\rho/\rho$ are calculated by applying the ideal gas law,
\beq
  \rho = \frac{M_\mathrm{c} P}{RT}\,,
  \qquad
  \frac{\Delta\rho}{\rho}
  = \frac{(M_\mathrm{c}-M_\mathrm{w})\,P_\mathrm{sat}}
         {M_\mathrm{c}P-(M_\mathrm{c}-M_\mathrm{w})\,P_\mathrm{sat}}\,,
  \label{eq_evap_in_term2}
\eeq
where $R$ is the universal gas constant. Parameter values are
given in Table~\ref{tab_para}.

\begin{table}[htb]
  \centering
  \begin{tabular}{ll} \hline
  Quantity & Value\rule{0em}{2.5ex}\\ \hline
  Gravity acceleration, $g$\rule{0em}{2.5ex} &
  $3.72\,\mathrm{m\,s^{-2}}$ \\
  Diffusion coefficient, $D$ &
  $1.4\times{}10^{-3}\,\mathrm{m^2\,s^{-1}}$ \\
  Kinematic viscosity of CO$_2$, $\nu$ &
  $6.93\times{}10^{-4}\,\mathrm{m^2\,s^{-1}}$ \\
  Universal gas constant, $R$ &
  $8.314\,\mathrm{J\,mol^{-1}\,K^{-1}}$ \\
  Molar mass of water, $M_\mathrm{w}$ &
  $1.802\times{}10^{-2}\,\mathrm{kg\,mol^{-1}}$ \\
  Molar mass of CO$_2$, $M_\mathrm{c}$ &
  $4.401\times{}10^{-2}\,\mathrm{kg\,mol^{-1}}$ \\ \hline
  \end{tabular}
  \caption{Physical parameters for the evaporation model of MAIC-2.}
  \label{tab_para}
\end{table}

\citet{sears_moore_05} state that
the evaporation rate of ice is probably about half that of
liquid water. In addition, any significant evaporation of the
dusty ice of the PLDs will lead to an enrichment of dust at the
surface, thus producing an insulating layer which hampers
further evaporation. These effects can be accounted for by
setting the evaporation factor $E_0$ in Eq.~(\ref{eq_evap_in})
to a value less than unity, and we will use a standard value of
$E_0=0.1$ (this choice will be detailed below in
Sect.~\ref{ssect_sim_transient}).

\begin{figure}[htb]
  \centering
  \includegraphics[width=80mm]{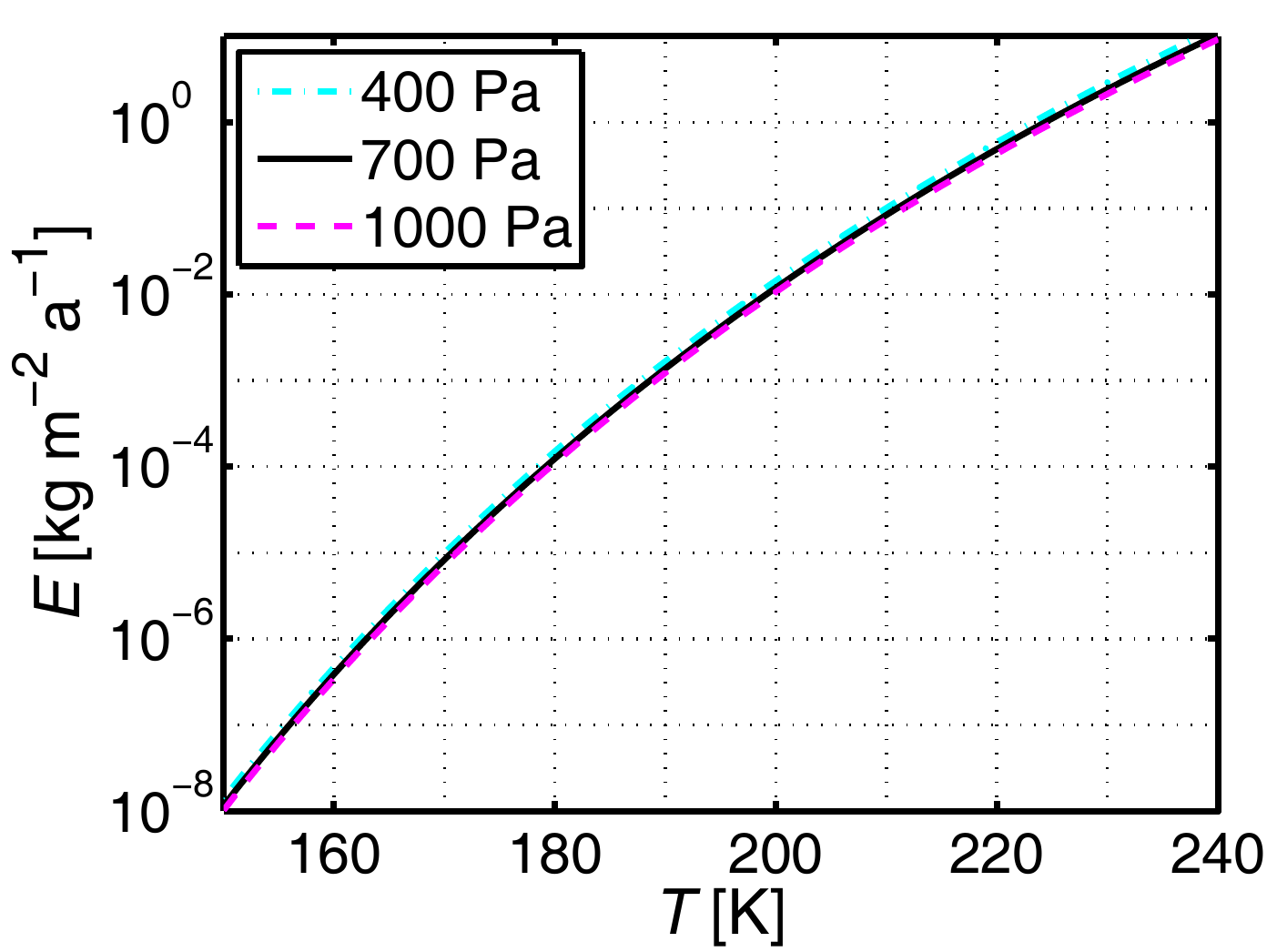}
  \par\vspace*{-1ex}\par
  \caption{Dependence of the evaporation rate $E$ on the surface
  temperature $T$ for surface pressures $P=400$, 700 and
  $1000\,\mathrm{Pa}$, and $E_0=0.1$.}
  \label{fig_evap}
\end{figure}

\begin{figure}[htb]
  \centering
  \includegraphics[width=80mm]{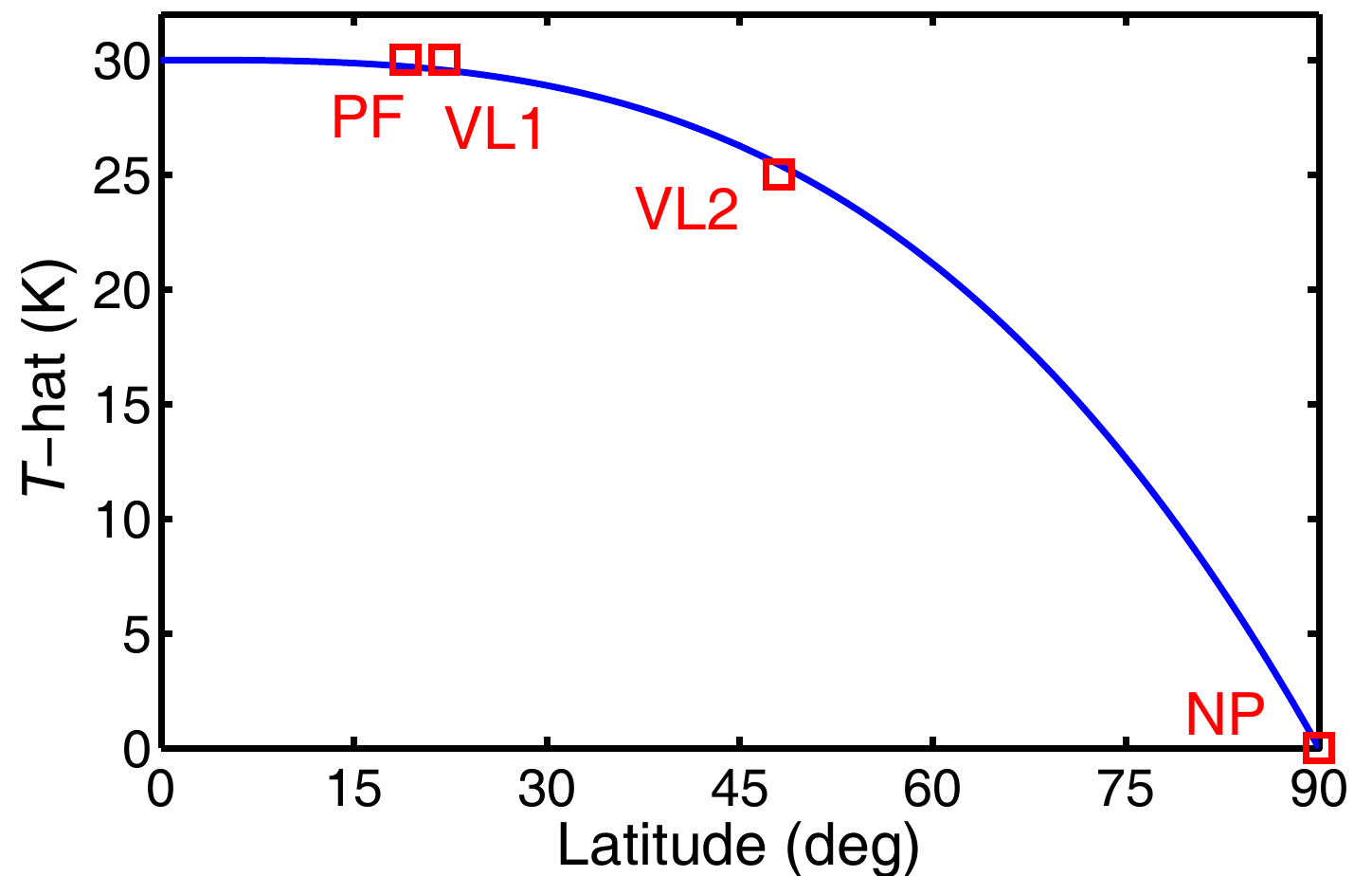}
  \par\vspace*{-1ex}\par
  \caption{Amplitude $\hat{T}$ of the daily cycle of the surface
  temperature according to Eq.~(\ref{eq_temp_daily_cycle_2}).
  Comparison with the data for Mars Pathfinder (PF), Viking Lander 1
  (VL1) and Viking Lander 2 (VL2) \citep{tillman_01} as well as the
  north pole (NP).}
  \label{fig_temp_daily_cycle}
\end{figure}

The dependence of evaporation on surface temperature and
pressure is illustrated in Fig.~\ref{fig_evap}. The temperature
dependence is evidently very strong, while the pressure
dependence is weak. Owing to the strong temperature dependence
and the short reaction time of evaporation on changing
conditions, it is not sufficient to calculate evaporation rates
on the basis of daily mean temperatures. Therefore, we
parameterise the daily cycle $T_\mathrm{DC}$ as follows,
\beq
   T_\mathrm{DC}(\varphi,t)
   = T(\varphi,t)
     + \hat{T}(\varphi)
       \,\cos\Big(\frac{2\pi{}t}{1\,\mbox{sol}}  \Big)\,,
  \label{eq_temp_daily_cycle_1}
\eeq
with the amplitude
\beq
   \hat{T}(\varphi)
   = \hat{T}_\mathrm{EQ}
      \left[ 1 - \Big(\frac{|\varphi|}{90^\circ}\Big)^3 \right]\,.
  \label{eq_temp_daily_cycle_2}
\eeq
The amplitude at the equator is set to
$\hat{T}_\mathrm{EQ}=30\,\mathrm{K}$. This choice provides a
good agreement with the amplitudes measured by the surface
missions Mars Pathfinder ($19\degN$,
$\hat{T}\sim{}30\,\mathrm{K}$), Viking Lander 1 ($22\degN$,
$\hat{T}\sim{}30\,\mathrm{K}$) and Viking Lander 2 ($48\degN$,
$\hat{T}\sim{}25\,\mathrm{K}$) \citep{tillman_01} as well as
the requirement $\hat{T}=0\,\mathrm{K}$ for the poles
(Fig.~\ref{fig_temp_daily_cycle}).

For high temperatures (e.g., $T_\mathrm{DC}>272.8\,\mathrm{K}$
for $P=700\,\mathrm{Pa}$), due to the increasing saturation
pressure $P_\mathrm{sat}$, the term $\Delta\rho/\rho$ of
Eq.~(\ref{eq_evap_in_term2}) becomes larger than unity, goes
through a singularity and then becomes negative. This means
that $\Delta\rho/\rho$ loses its physical meaning. In that
case, we correct the problem by resetting $\Delta\rho/\rho$ to
unity.

The above equation (\ref{eq_evap_in}) is valid for an ice cap
in contact with the atmosphere. By contrast, for the case of
ground ice, we assume that the evaporation rate is reduced with
increasing thickness $H_\mathrm{reg}$ of the ice-free regolith
layer (which separates the atmosphere from the ground ice),
\beq
  E \rightarrow
  E \times \exp\left(-\frac{H_\mathrm{reg}}{\gamma_\mathrm{reg}}\right)\,,
  \label{eq_evap_reg}
\eeq
where $\gamma_\mathrm{reg}$ is the regolith-insulation
coefficient, chosen as $\gamma_\mathrm{reg}=0.1\,\mathrm{m}$.

\subsection{Condensation}
\label{ssect_cond}

The water content $\omega$ in the atmosphere is expressed as an
area mass density in units of $\mathrm{kg\,m^{-2}}$. Multiplied
with the gravity acceleration $g$, this becomes equivalent to
the partial pressure of water vapour at the surface. Thus we
compare this pressure to the water vapor saturation pressure
$P_{\mathrm{sat}}$ and assume that all excess water condenses
instantly,
\beq
  \begin{array}{ll}
    \mbox{If } g\omega > P_{\mathrm{sat}}(T):
    & \mbox{excess water }
    [\omega - P_{\mathrm{sat}}(T)/g]
    \\
    & \mbox{determines condensation rate } C\,,
    \\[1ex]
    \mbox{else}:
    & C = 0\,.
  \end{array}
  \label{eq_cond_excess}
\eeq
Note that this is a very simplistic approach because in reality
condensation takes place higher in the atmosphere where the
temperature may differ considerably from the surface
temperature.

\subsection{Transport}
\label{ssect_transport}

As mentioned above, the water content $\omega$ in the
atmosphere is an area mass density. Since the evaporation $E$
(cf.~Sect.~\ref{ssect_evap}) and condensation $C$
(cf.~Sect.~\ref{ssect_cond}) are expressed as mass fluxes in
units of $\mathrm{kg\,m^{-2}\,s^{-1}}$, its evolution is
governed by
\beq
  \frac{\partial \omega}{\partial t}
  = - \nabla\cdot\vecbu{F} + E - C\,,
  \label{eq_balance}
\eeq
where $\vecbu{F}$ is the horizontal water flux in units of
$\mathrm{kg\,m^{-1}\,s^{-1}}$.

We approximate the atmospheric water transport by instantaneous
mixing (on a time scale of several sols). This can formally be
obtained by assuming a diffusive flux,
\beq
  \vecbu{F} = - K \, \nabla \omega\,,
  \label{eq_fluxes}
\eeq
with the limit of infinite diffusivity, $K\rightarrow\infty$.

The MAIC version with the LIT scheme of Sect.~\ref{ssect_temp}
and the surface mass balance of water ice that results from
Sects.~\ref{ssect_pres_sat}--\ref{ssect_transport} is referred to
as ``MAIC-2''.

\subsection{Ice evolution}
\label{ssect_ice}

With the condensation $C$ and the evaporation $E$, the net mass
balance $a_\mathrm{net}$ of the ice caps, expressed in units of
$\mathrm{m\,s^{-1}}$ ice equivalent, is
\beq
  a_\mathrm{net} = \frac{C-E}{\rho_\mathrm{ice}}\,,
\eeq
where $\rho_\mathrm{ice}=910\,\mathrm{kg\,m^{-3}}$ is the
density of ice. For a static model (glacial flow neglected),
the evolution of the ice thickness, $H$, is then simply
\beq
  \pabl{H}{t} = a_\mathrm{net}\,.
  \label{eq_thickness}
\eeq
Note that we allow for negative ice thicknesses ($H<0$). Such a
situation is interpreted as ground ice under an ice-free
regolith layer of thickness $H_\mathrm{reg}=|H|$. The thickness
of the ground ice layer itself is undefined.

The validity of the assumption of negligible glacial flow is
debatable. On the one hand, modelled ice flow speeds on Mars
are generally slow, even during periods of high obliquity
\citep{greve_etal_04a, greve_mahajan_05}. One the other hand,
locally enhanced glacial flow may occur near chasmata and
troughs of the PLDs \citep{hvidberg_03, greve_08a}, and
\citet{winebrenner_etal_08} argue that the overall topography
of Gemina Lingula (also known as Titania Lobe), the lobe of the
northern PLDs to the south of Chasma Boreale, was likely shaped
by past glacial flow. In this study, we will stick to the
simple, static ice model, but consider the inclusion of glacial
flow for future work.

\section{Simulations}
\label{sect_sim}

We will now discuss the application of MAIC-2 to two different
sets of scenarios. The first set is of ``academic'' nature with
orbital parameters kept constant over time, whereas the second
one employs a realistic, time-dependent forcing over the last
10 million years. In order to carry out these simulations, a
finite-difference/finite-volume discretisation of the model
equations of MAIC-2 has been derived (see
Appendix~\ref{sect_discrete} for details) and coded in the
Fortran program maic2.F90 (available as free software at
http://maic2.greveweb.net). Instantaneous mixing of water
vapour in the atmosphere is assumed (diffusivity
$K\rightarrow\infty$), a time step of
$\Delta{}t=0.02\,\mathrm{a}$ ($\approx{}7\,\mathrm{sols}$) and
an equidistant grid spacing of $\Delta\varphi=1^\circ$ are
chosen (the formulation in Appendix~\ref{sect_discrete} allows
also for non-equidistant grid spacings), and the initial ice
distribution is a layer of 19~m thickness on the entire surface
of Mars. The latter value accounts for the ice inventory of the
present-day PLDs, $\sim{}1.14\times{}10^6\,\mathrm{km^3}$ in
the north \citep{grima_etal_09} and
$\sim{}1.6\times{}10^6\,\mathrm{km^3}$ in the south
\citep{plaut_etal_07}.

\subsection{Constant orbital parameters}
\label{ssect_sim_constant}

Simulations \#1--4 have been run over 10 million years with
constant orbital parameters as follows:
\begin{itemize}
\item Obliquities: $\phi=15^\circ$ (\#1), $25.2^\circ$
    (present-day value, \#2), $35^\circ$ (\#3) and
    $45^\circ$ (\#4).
\item Eccentricity: $\eps=0.093$ (present-day value,
    \#1--4).
\item Solar longitude of perihelion:
    $L_\mathrm{s,p}=251.0^\circ$ (present-day value,
    \#1--4).
\end{itemize}
The evaporation factor in Eq.~(\ref{eq_evap_in}) is set to the
standard value $E_0=0.1$ for all four simulations.

\begin{figure}[htb]
  \centering
  \includegraphics[width=110mm]{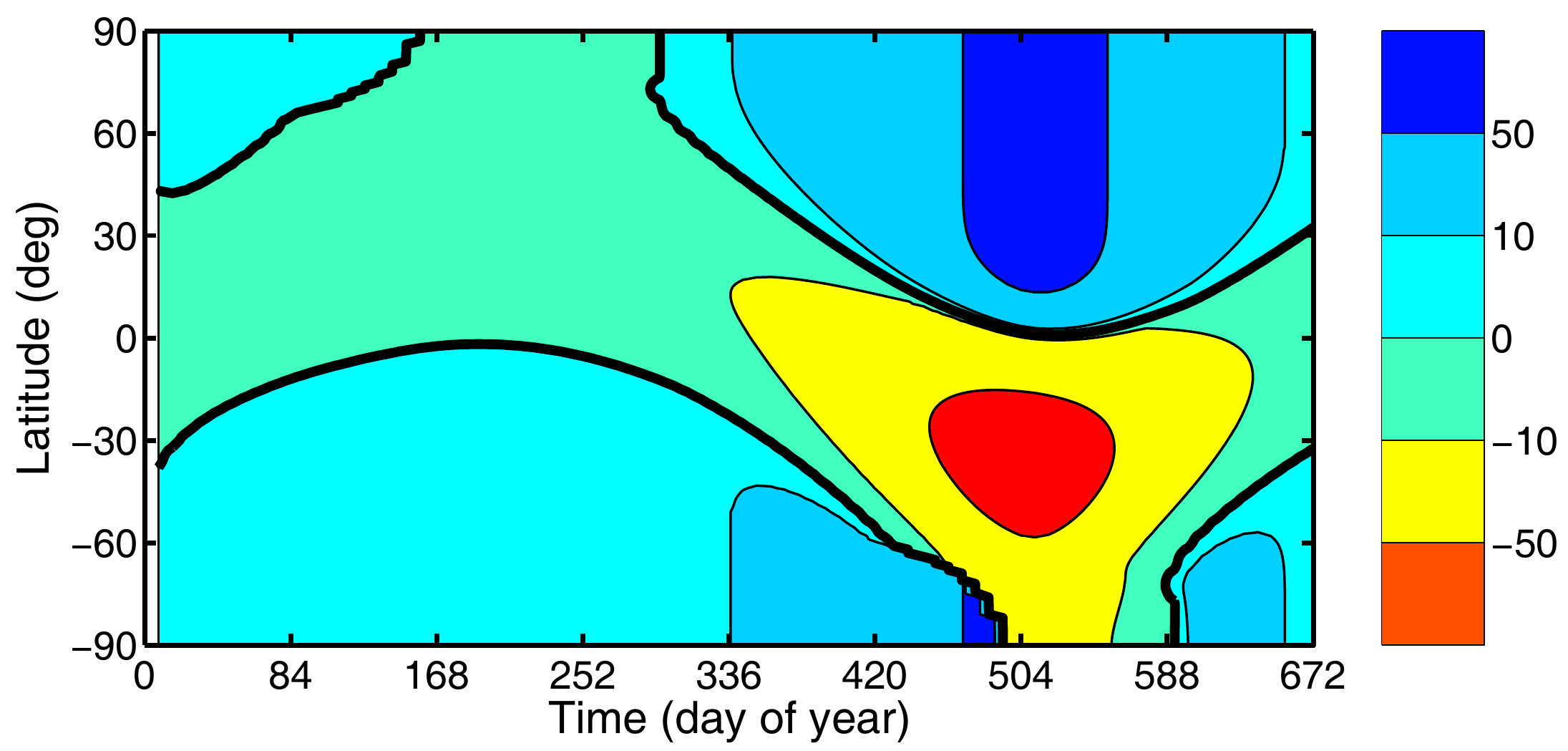}
  \par\vspace*{-1ex}\par
  \caption{Simulation \#2:
  Net surface mass balance (in $\mathrm{mm\,ice\,equiv.\,a^{-1}}$)
  in the first Martian year. The thick contour shows the equilibrium
  line (zero mass balance).}
  \label{fig_run_c01a_2}
\end{figure}

The resulting net mass balance of water ice in the first
Martian year of simulation \#2 (all parameters at their
present-day values) is shown in Fig.~\ref{fig_run_c01a_2}. The
distribution resembles that of the surface temperature
(Fig.~\ref{fig_lit_temp}a). The seasonal CO$_2$ caps are
efficient cold traps for atmospheric water vapour, which leads
to strongly positive mass balances in the Martian polar regions
for most of the year. By contrast, in the lower latitudes
negative mass balances prevail, so that the initial ice layer
of constant thickness is redistributed towards the poles.

\begin{figure}[htb]
  \centering
  \includegraphics[width=0.98\textwidth]{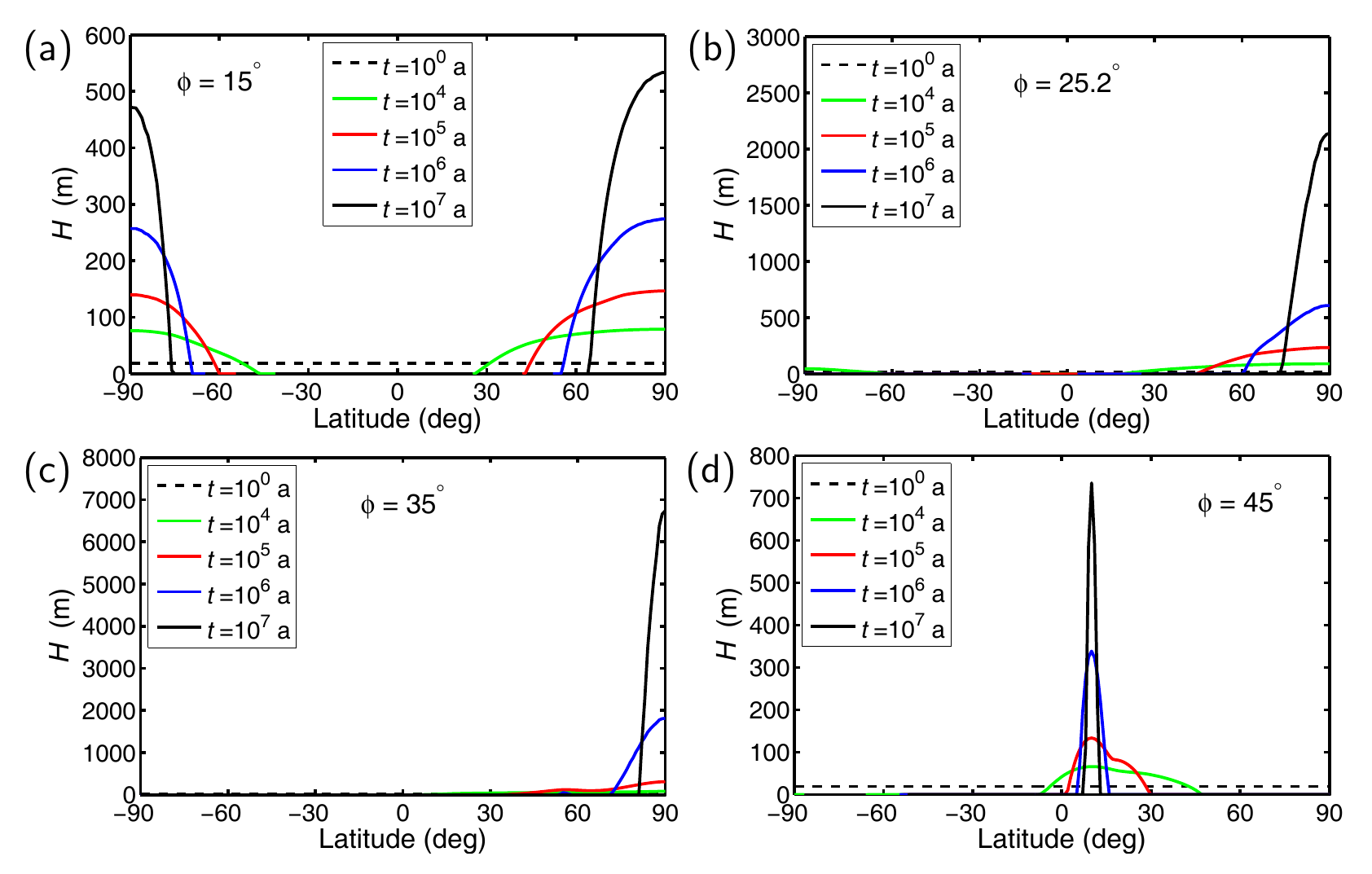}
  \par\vspace*{-1ex}\par
  \caption{(a) Simulation \#1 ($\phi=15^\circ$),
  (b) \#2 ($\phi=25.2^\circ$),
  (c) \#3 ($\phi=35^\circ$) and
  (d) \#4 ($\phi=45^\circ$):
  Evolution of the ice thickness $H$.
  Note the different scales of the $H$-axes.}
  \label{fig_runs_c01_c02_c03_c04}
\end{figure}

The evolution of the ice thickness over the entire simulation
time of $10^7$ years for all simulations is depicted in
Fig.~\ref{fig_runs_c01_c02_c03_c04}. The obliquity of
simulation \#1 ($\phi=15^\circ$) is approximately equal to the
minimum value which occurred during the last 4~Ma
(Fig.~\ref{fig_obliq}). The resulting evolution of the ice
thickness is shown in Fig.~\ref{fig_runs_c01_c02_c03_c04}a. The
simulation produces bipolar ice deposits, somewhat more
pronounced in the north than in the south. After $10^7$ years,
the simulated polar deposits reach maximum thicknesses of
$\sim{}500\,\mathrm{m}$, which is about a factor 5 thinner than
the observed PLDs at present.

For simulation \#2 (Fig.~\ref{fig_runs_c01_c02_c03_c04}b), it
is striking that the redistribution of ice towards the poles is
strongly antisymmetric. Until $10^4$ years, MAIC-2 produces
more pronounced ice deposits in the northern hemisphere and
less pronounced ones in the southern hemisphere. At $10^5$
years and later, the ice migrates entirely to the northern
hemisphere and concentrates around the north pole. In fact, the
simulated north polar deposits at $10^7$ years resemble the
currently existing PLDs in extent and thickness (see below,
Fig.~\ref{fig_run_t06_evol}b), while the simulated south polar
region is ice-free.

The reason for this behaviour is the hemispheric asymmetry of
the daily mean surface temperature (Fig.~\ref{fig_lit_temp}).
As a consequence of the closer proximity of Mars to the Sun
during southern summer, the southern summer is distinctly
warmer than the northern summer. This leads to large
evaporation rates during southern summer and thus a large
amount of water vapour in the atmosphere, which is trapped
preferably in the winter-cold high northern latitudes.
Conversely, the northern summer is less warm, evaporation rates
are lower, and therefore the potential for water ice
accumulation in the south polar region is much smaller (see
also Fig.~\ref{fig_run_c01a_2}). This holds also for simulation
\#1; however, the effect is less pronounced as a result of the
weaker seasonal cycle due to the lower obliquity. Hence the
hemispheric asymmetry is much weaker for simulation \#1 than
for simulation \#2.

For simulation \#3, the obliquity ($\phi=35^\circ$) is
approximately equal to the maximum value during the last 4~Ma.
Figure~\ref{fig_runs_c01_c02_c03_c04}c displays the resulting
evolution of the ice thickness. The result is similar to that
of simulation \#2; however, the concentration of ice around the
north pole is more extreme. The simulated north polar deposits
at $10^7$ years are as thick as 6.7~km, almost 2.5 times
thicker than their observed present-day counterparts.

The obliquity of simulation \#4 ($\phi=45^\circ$) was reached
in several maxima during the period of high average obliquity
prior to 5~Ma ago. This makes the seasons very extreme. Like in
simulations \#1--3, the ice is preferentially deposited in the
northern hemisphere (Fig.~\ref{fig_runs_c01_c02_c03_c04}d) due
to the warmer southern summers. However, now the poles receive
substantial insolation during the respective summer season, so
that the northern hemispheric ice deposits are not thickest at
the north pole any more. Instead, ice deposition is favoured in
the low latitudes, and beyond $10^4$ years simulation time the
deposits develop a thickness maximum as far south as
$\!\!~\sim\!{}10\degN$.

\subsection{Evolution over the last ten million years}
\label{ssect_sim_transient}

Simulations \#5--8 attempt at providing realistic,
time-dependent scenarios over the last millions of years. To
this end, they have been run from 10~Ma ago until today, driven
by the history of orbital parameters (obliquity, eccentricity,
solar longitude of perihelion) by \citet{laskar_etal_04}. The
evaporation factor in Eq.~(\ref{eq_evap_in}) is set to the
values $E_0=0.05$ (\#5), 0.1 (standard value, \#6), 0.2 (\#7)
and 0.3 (\#8).

\begin{table}[htb]
  \centering
  \begin{tabular}{clccccc} \hline
  Sim.\rule{0em}{2.5ex}
  & $E_0$
  & $V_\mathrm{NPLD}\;[\mathrm{km^3}]$
  & $V_\mathrm{SPLD}\;[\mathrm{km^3}]$
  & $H_\mathrm{NP}\;[\mathrm{m}]$
  & $H_\mathrm{SP}\;[\mathrm{m}]$
  & $J$ \\ \hline
  \#5\rule{0em}{2.5ex}
  & 0.05
  & $1.24\times{}10^6$
  & $1.58\times{}10^6$
  & 1889
  & 2114
  & 9.30 \\
  \#6
  & 0.1
  & $1.16\times{}10^6$
  & $1.65\times{}10^6$
  & 2404
  & 2732
  & 2.11 \\
  \#7
  & 0.2
  & $1.02\times{}10^6$
  & $1.78\times{}10^6$
  & 2577
  & 3170
  & 4.96\\
  \#8
  & 0.3
  & $0.94\times{}10^6$
  & $1.86\times{}10^6$
  & 2431
  & 3751
  & 12.05\hspace*{0.5em}\\ \hline
  Obs.\rule{0em}{2.5ex}
  & ---
  & $1.14\times{}10^6$
  & $1.6\hspace*{0.5em}\times{}10^6$
  & 2773
  & 2285
  & ---\\ \hline
  \end{tabular}
  \caption{Simulations \#5--8: Evaporation factor $E_0$,
  present-day volume of the north and south polar layered
  deposits ($V_\mathrm{NPLD}$, $V_\mathrm{SPLD}$),
  present-day ice thickness at the north and south pole
  ($H_\mathrm{NP}$, $H_\mathrm{SP}$).
  The last row shows the observed volumes
  \citep{plaut_etal_07, grima_etal_09}
  and ice thicknesses \citep[][see also the caption of
  Fig.~\ref{fig_run_t06_evol}]{zuber_etal_98, smith_etal_99a,
  greve_etal_04a}.
  The misfits $J$ have been computed according to Eq.~(\ref{eq_misfit}).}
  \label{tab_transient_runs}
\end{table}

For the present ($t=0$), all simulations produce bipolar ice
deposits. An overview of the results is given in
Table~\ref{tab_transient_runs}. It shows that, with increasing
evaporation factor $E_0$, the volume of the northern deposits
decreases, while the volume of the southern deposits increases.
This holds essentially also for the ice thicknesses at the
poles. The only exception is the north polar thickness of
simulation \#8 compared to \#7, which decreases by
$\sim\!{}6\%$ even though the ice volume increases by
$\sim\!{}4\%$. These distinctive trends allow to identify the
most suitable value of $E_0$. To this end, we define the misfit
$J$ as follows,
\beqa
  J &=&
  \left(\frac{V_\mathrm{NPLD}-V_\mathrm{NPLD,obs}}
             {\sigma_{V_\mathrm{NPDL}}}\right)^2
  +
  \left(\frac{V_\mathrm{SPLD}-V_\mathrm{SPLD,obs}}
             {\sigma_{V_\mathrm{SPDL}}}\right)^2
  \nl &&
  +
  \left(\frac{H_\mathrm{NP}-H_\mathrm{NP,obs}}
             {\sigma_{H_\mathrm{NP}}}\right)^2
  +
  \left(\frac{H_\mathrm{SP}-H_\mathrm{SP,obs}}
             {\sigma_{H_\mathrm{SP}}}\right)^2\,.
  \label{eq_misfit}
\eeqa
The standard deviations are introduced to make the various
contributions to $J$ commensurate. They are computed from the
four respective values of simulations \#5--8,
\beq
  \begin{array}{rclrcl}
    \sigma_{V_\mathrm{NPDL}} &=& 0.134\times{}10^6\,\mathrm{km^3}\,,
    \\
    \sigma_{V_\mathrm{SPLD}} &=& 0.126\times{}10^6\,\mathrm{km^3}\,,
    \\
    \sigma_{H_\mathrm{NP}}   &=& 301\,\mathrm{m}\,,
    \\
    \sigma_{H_\mathrm{SP}}   &=& 692\,\mathrm{m}\,.
  \end{array}
  \label{eq_misfit_sigma}
\eeq
The resulting misfits $J$ are listed in the last column of
Table~\ref{tab_transient_runs}. Simulation \#6 with $E_0=0.1$
produces clearly the best agreement (and thus $E_0=0.1$ is used
as standard value throughout this study). The errors of the
volumes are as small as 1.6\% for the NPLD and 3.2\% for the
SPLD, the ice thickness is 13.3\% too small at the north pole
and 19.6\% too large at the south pole. Given the simplicity of
the MAIC-2 model, this is a remarkably good overall agreement.

\begin{figure}[htb]
  \centering
  \includegraphics[width=80mm]{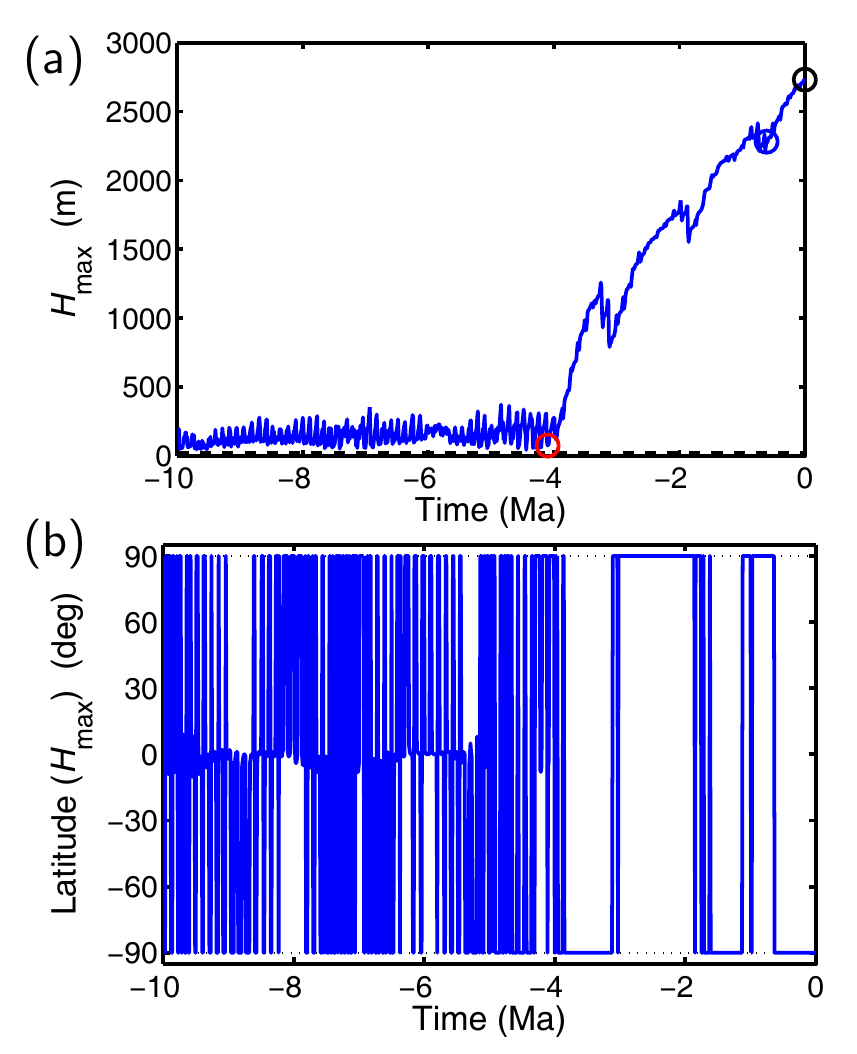}
  \par\vspace*{-1ex}\par
  \caption{Simulation \#6:
  (a) Maximum ice thickness (the circle marks correspond to the time
  slices shown in Fig.~\ref{fig_run_t06_evol}).
  (b) Latitude of maximum ice thickness.}
  \label{fig_run_t06}
\end{figure}

In the following, the best-fit simulation \#6 will be
discussed. The maximum ice thickness and its position on the
planet are shown in Fig.~\ref{fig_run_t06}. The simulation
produces a mobile glaciation with two distinctly different
stages. \emph{Stage~1}, the period of high average obliquity
from 10 until 4~Ma ago, is characterised by ice thicknesses
less than 400~m and rapid changes of the position where the
maximum thickness occurs between all latitudes. By contrast,
during \emph{stage~2}, the period of low average obliquity from
4~Ma ago until today, the position of maximum thickness changes
much less rapidly and flip-flops between the poles only (47\%
of the time at the north pole, 53\% of the time at the south
pole). The polar ice deposits grow almost monotonically to
their present-day thicknesses, only interrupted by moderate
shrinking around $\sim\!{}3.2$, 1.9 and 0.7~Ma ago when maximum
amplitudes of the main obliquity cycle of 125~ka occurred (see
also Fig.~\ref{fig_obliq}).

\begin{figure}[htb]
  \centering
  \includegraphics[width=0.98\textwidth]{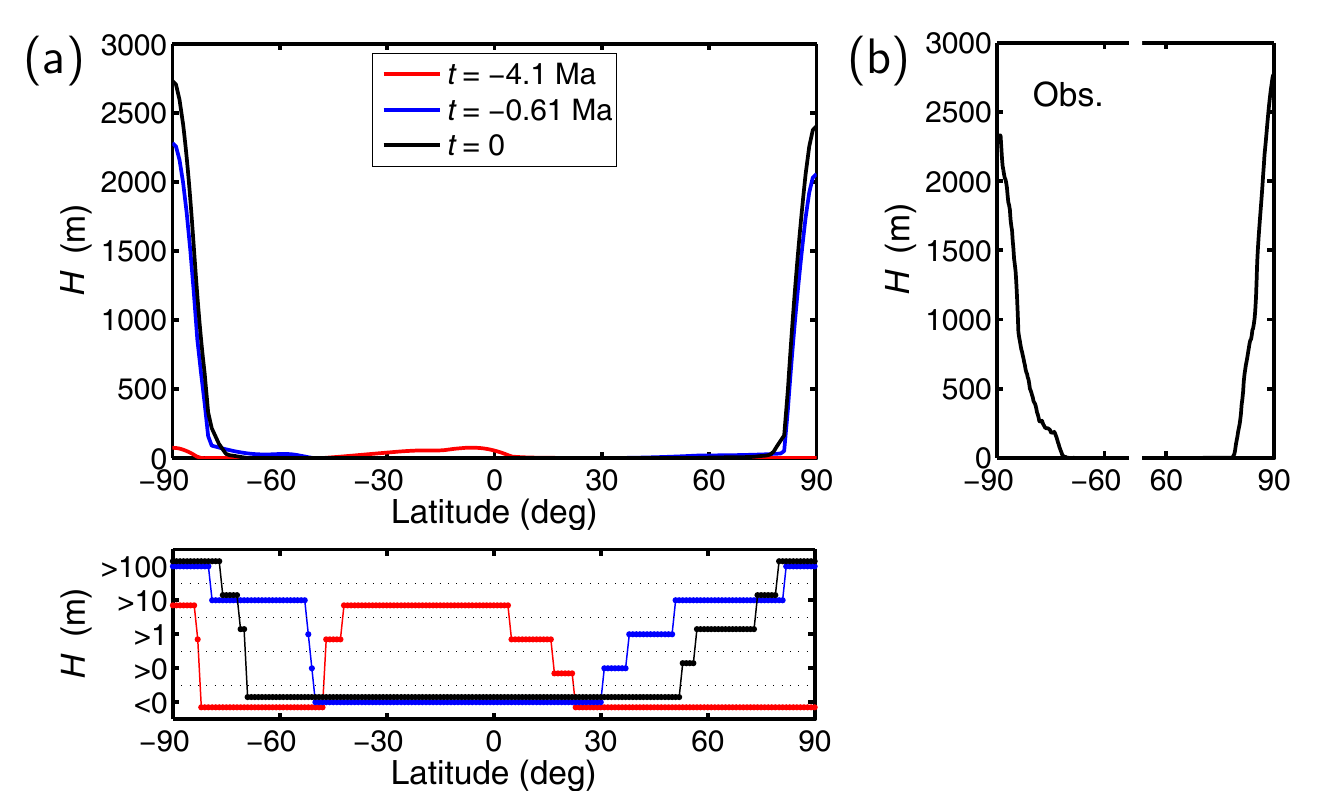}
  \par\vspace*{-1ex}\par
  \caption{(a) Simulation \#6:
  Ice thickness $H$ for three selected time slices (which correspond
  to the marks in Fig.~\ref{fig_run_t06}a.)
  The bottom panel shows thickness \emph{classes} in order to highlight
  deposits of thin ice not visible in the upper panel.
  The class ``$\,<\!0\,$'' means ground ice (see Sect.~\ref{ssect_ice}).
  (b) Observational data of the ice thickness
  of the present-day PLDs. They have been obtained by subtracting the
  MOLA surface topography \citep{zuber_etal_98, smith_etal_99a} from
  the basal topography computed by a smooth extrapolation of the
  ice-free ground \citep{greve_etal_04a} and subsequent zonal averaging.}
  \label{fig_run_t06_evol}
\end{figure}

In order to illustrate this behaviour in more detail,
Fig.~\ref{fig_run_t06_evol} depicts the distribution of the
simulated ice thickness for three selected time slices. The
first time slice, 4.1~Ma ago (near the end of stage~1), is
characterised by a low maximum ice thickness (74.4~m) which
occurs in the low southern latitudes (at $6\degS$) and a
continuous glaciation from the mid southern ($47\degS$) to the
low northern ($22\degN$) latitudes. In addition, small south
polar deposits extend from the pole to $83\degS$, whereas the
north polar region is entirely ice-free.

The second time slice, 0.61~Ma ago (in stage~2, towards the end
of the most recent period of large obliquity amplitudes), shows
polar ice deposits only moderately smaller than their
present-day counterparts. However, the striking difference
compared to the present is the existence of continuous, at
least metres-thick ice deposits equatorward to $38\degN$ and
$52\degS$, respectively. These mid-latitude deposits are very
mobile; merely 20~ka earlier (0.63~Ma ago), they are entirely
absent in the northern hemisphere while extending equatorward
to $24\degS$ in the southern hemisphere (not shown).

The third time slice is the present ($t=0$), with an obliquity
of $25.2^\circ$ following a $\sim{}0.3\;\mathrm{Ma}$ period
with only small changes (within less than $5^\circ$). As
already discussed above, for the present, the simulation
predicts bipolar ice deposits which match the observed volumes
within $\sim\!{}3\%$ and the ice thicknesses at the poles
within $\sim\!{}20\%$. The bulk of the deposits with
thicknesses $\ge{}100\,\mathrm{m}$ extend to $80\degN$ and
$77\degS$, which also agrees well with the observed values of
$80\degN$ and $73\degS$, respectively. Metres-thick deposits
follow equatorward to $57\degN$ and $70\degS$. This is still
compatible with the reality in the south, while it contradicts
the reality in the north where such deposits are not observed.

\section{Discussion and conclusion}

The simulations with constant orbital parameters presented in
Sect.~\ref{ssect_sim_constant} confirm the intuitive idea that
low obliquities favour deposition of water ice in high
latitudes and vice versa. An interesting additional finding is
that the polar ice deposits for relatively low obliquities can
either occur at one pole only or at both poles, depending on
the asymmetry of the seasons in the two hemispheres.

The more realistic best-fit simulation \#6 of
Sect.~\ref{ssect_sim_transient} with time-dependent orbital
forcing from 10~Ma ago until today produces a very good
agreement with the observed present-day PLDs. It predicts a
mobile glaciation with two distinct stages. During stage~1,
from 10 to 4~Ma ago, ice thicknesses never extend 400~m, and
ice is readily exchanged between all latitudes. This exchange
is mainly controlled by obliquity, polar deposits being again
favoured by relatively low obliquities and lower-latitude
deposits by peak obliquities. During stage~2, from 4~Ma ago
until today, the north and south polar ice deposits grow
essentially monotonically and reach their maximum thicknesses
at the present. In particular during the three periods of large
obliquity amplitudes around $\sim\!{}3.2$, 1.9 and 0.7~Ma ago,
the polar deposits continue in thin, very mobile ice deposits
which extend far into the mid latitudes at times. The latter
result agrees with the findings by \citet{head_etal_03} who
report evidence for ``ice ages'' on Mars during the period from
about 2.1 to 0.4~Ma ago when the obliquity regularly exceeded
$30^\circ$. According to the authors, the conditions during
this period favoured the deposition of metres-thick, dusty,
water-ice-rich material down to latitudes of $\sim{}30^\circ$
in both hemispheres.

A limitation of the results of this study must be noted. The
estimated surface ages of the northern (at most
$0.1\;\mathrm{Ma}$) and southern PLDs (about $10\;\mathrm{Ma}$)
by \citet{herkenhoff_plaut_00}, which are based on crater
statistics, are consistent with the simulated growth of ice
deposits during the last 4~Ma for the north, but not for the
south polar region (see Fig.~\ref{fig_run_t06_evol}). This may
be related to the fact that most of the southern PLDs are
covered by dust, whereas the water ice of their northern
counterpart is exposed to the atmosphere. Consequently, at
least for the present-day situation, the northern PLDs can
readily exchange water with the atmosphere, whereas the
exchange is blocked to an unknown extent for the southern PLDs.
This problem requires further attention. Nevertheless, we
conclude that the model MAIC-2 is a very useful tool which,
despite its simplicity, can provide substantial insight in the
evolution of the Martian surface glaciation over climatological
time scales.

\begin{small}

\section*{Acknowledgements}

The efforts of the scientific editor and reviewers (anonymous)
are gratefully acknowledged. This work was partly carried out
within the project ``Evolution and dynamics of the Martian
polar ice caps over climatic cycles'' supported by the Research
Fund of the Institute of Low Temperature Science, Sapporo,
Japan.

\begin{appendix}

\section{Discrete formulation}
\label{sect_discrete}

\subsection{Numerical grid}
\label{ssect_grid}

MAIC-2 is a spatially one-dimensional model which features a
dependence on latitude only. The grid points are located at
monotonically increasing latitudes,
\beq
   \varphi_l\,, \quad l=0,\ldots,L\,,
\eeq
where $\varphi_0=-{\pi}/{2}$ (south pole) and
$\varphi_L={\pi}/{2}$ (north pole). The generally
non-equidistant spacing between subsequent grid points is
\beq
   \Delta\varphi_l = \varphi_l - \varphi_{l-1}\,, \quad l=1,\ldots,L\,.
\eeq
Further, we define the latitudes in between grid points (at
cell boundaries) as
\beq
   \varphi_{l\pm{}1/2} = \frac{\varphi_l+\varphi_{l\pm{}1}}{2}\,.
\eeq
Time is discretised uniformly by
\beq
  t^n = t^0 + n\,\Delta{}t\,, \quad n=0,\ldots,N\,,
\eeq
where $t^0$ is the initial time, $\Delta{}t$ the time step and
$t^N=t^0+N\,\Delta{}t$ the final time.

\subsection{Surface temperature, evaporation, condensation}
\label{ssect_diff_temp_evap_cond}

Numerical evaluation of the LIT
[Eqs.~(\ref{eq_rad_balance})--(\ref{eq_w_cond_evap})] and
evaporation [Eqs.~(\ref{eq_evap_in})--(\ref{eq_evap_reg})]
schemes is essentially straightforward and need not be
detailed. The daily cycle of the surface temperature
[Eq.~(\ref{eq_temp_daily_cycle_1})] for computing evaporation
rates is sampled with the sub-daily time step
$\Delta{}t_\mathrm{DC}=\frac{1}{8}\,\mathrm{sol}$. As for
condensation, the discrete version of the condition
(\ref{eq_cond_excess}) is
\beq
  C_l^n = \left\{
  \begin{array}{cl}
    \bigfrac{1}{\Delta t} \,
    \left( \omega_l^n - \bigfrac{P_{\mathrm{sat}}(T_l^n)}{g} \right)\,,
    & \mbox{if } g\omega_l^n > P_{\mathrm{sat}}(T_l^n)\,,
    \\[3ex]
    0\,, & \mbox{otherwise}\,.
  \end{array} \right.
  \label{eq_cond_excess_disc}
\eeq

\subsection{Instantaneous mixing}
\label{ssect_inst_discrete}

As mentioned in Sect.~\ref{ssect_transport}, instantaneous
mixing of water vapour in the atmosphere can be described as
the limit of infinite diffusivity, $K\rightarrow\infty$. A
finite-volume discretisation of the diffusion equation with
finite diffusivity is described in an earlier version of this
paper (archived at arXiv:0903.2688v1 [physics.geo-ph]);
however, the limit $K\rightarrow\infty$ cannot be carried out
numerically with that scheme. Instead, a two-step procedure has
been devised for the case of instantaneous mixing.

In the first step, for any point of the model domain
($l=0,\ldots,L$), let us compute predictors of the water
content at the new time step, $\hat{\omega}_l^{n}$, by ignoring
the water transport,
\beq
  \frac{\hat{\omega}_l^{n}-\omega_l^{n-1}}{\Delta t} = E_l^n - C_l^n\,.
  \label{eq_diffusion_im_1}
\eeq
In the second step, the resulting total amount of water shall
be mixed over the entire planet. The total amount of water is
obtained by integrating over the grid cells and summing up,
\beqa
  \omega_\mathrm{tot}
  &=& 2\pi R^2 \hat{\omega}_0^{n} \int\limits_{\varphi_0}^{\varphi_{1/2}}\cos\varphi\,\D\varphi
      + \sum\limits_{l=1}^{L-1}
        2\pi R^2 \hat{\omega}_l^{n} \int\limits_{\varphi_{l-1/2}}^{\varphi_{l+1/2}}\cos\varphi\,\D\varphi
  \nl &&
      +\;2\pi R^2 \hat{\omega}_L^{n} \int\limits_{\varphi_{L-1/2}}^{\varphi_L}\cos\varphi\,\D\varphi
  \nl
  &=& 2\pi R^2\left\{
      \hat{\omega}_0^{n}\,(1+\sin\varphi_{1/2})
      + \sum\limits_{l=1}^{L-1} \hat{\omega}_l^{n}\,(\sin\varphi_{l+1/2}-\sin\varphi_{l-1/2})
      \right.
  \nl && \left.
      \hspace*{3em} +\;\hat{\omega}_L^{n}\,(1-\sin\varphi_{L-1/2}) \right\}\,,
\eeqa
where $R$ denotes the radius of the planet
($R=3396\;\mathrm{km}$). For all points $l=0,\ldots,L$, the new
water content follows by division by the surface of the planet,
\beqa
  \omega_l^{n} = \frac{\omega_\mathrm{tot}}{4\pi R^2}
  &=& \frac{1}{2} \left\{
      \hat{\omega}_0^{n}\,(1+\sin\varphi_{1/2})
      + \sum\limits_{l=1}^{L-1} \hat{\omega}_l^{n}\,(\sin\varphi_{l+1/2}-\sin\varphi_{l-1/2})
      \right.
  \nl && \left.
      \hspace*{1.4em} +\;\hat{\omega}_L^{n}\,(1-\sin\varphi_{L-1/2}) \right\}\,.
\eeqa

\subsection{Ice evolution}
\label{ssect_ice_discrete}

The discretisation of the ice-thickness equation
(\ref{eq_thickness}) is straightforward. By using an Euler
backward step for the time derivative, we obtain
\beq
  \frac{H_l^{n}-H_l^{n-1}}{\Delta t}
  = (a_\mathrm{net})_l^n
  = \frac{C_l^n-E_l^n}{\rho_\mathrm{ice}}\,.
\eeq

\end{appendix}



\begin{thebibliography}{30}
\providecommand{\natexlab}[1]{#1}
\providecommand{\url}[1]{\texttt{#1}}
\providecommand{\urlprefix}{URL }
\expandafter\ifx\csname urlstyle\endcsname\relax
  \providecommand{\doi}[1]{doi:\discretionary{}{}{}#1}\else
  \providecommand{\doi}{doi:\discretionary{}{}{}\begingroup
  \urlstyle{rm}\Url}\fi

\bibitem[{Armstrong et~al.(2004)Armstrong, Leovy and Quinn}]{armstrong_etal_04}
Armstrong, J.~C., C.~B. Leovy and T.~Quinn. 2004.
\newblock A {1~Gyr} climate model for {Mars:} new orbital statistics and the
  importance of seasonally resolved polar processes.
\newblock \emph{{Icarus}}, \textbf{171}~(2), 255--271.
\newblock \doi{10.1016/j.icarus.2004.05.007}.

\bibitem[{Forget et~al.(1999)Forget, Hourdin, Fournier, Hourdin, Talagrand,
  Collins, Lewis, Read and Huot}]{forget_etal_99}
Forget, F., F.~Hourdin, R.~Fournier, C.~Hourdin, O.~Talagrand, M.~Collins,
  S.~R. Lewis, P.~L. Read and J.-P. Huot. 1999.
\newblock Improved general circulation models of the {Martian} atmosphere from
  the surface to above 80~km.
\newblock \emph{{J. Geophys. Res.}}, \textbf{104}~(E10), 24155--24175.

\bibitem[{Gierasch and Goody(1968)}]{gierasch_goody_68}
Gierasch, P. and R.~Goody. 1968.
\newblock A study of the thermal and dynamical structure of the martian lower
  atmosphere.
\newblock \emph{{Planet. Space Sci.}}, \textbf{16}~(5), 615--646.

\bibitem[{Greve(2008)}]{greve_08a}
Greve, R. 2008.
\newblock Scenarios for the formation of {Chasma Boreale, Mars}.
\newblock \emph{{Icarus}}, \textbf{196}~(2), 359--367.
\newblock \doi{10.1016/j.icarus.2007.10.020}.

\bibitem[{Greve and Mahajan(2005)}]{greve_mahajan_05}
Greve, R. and R.~A. Mahajan. 2005.
\newblock Influence of ice rheology and dust content on the dynamics of the
  north-polar cap of {Mars}.
\newblock \emph{{Icarus}}, \textbf{174}~(2), 475--485.
\newblock \doi{10.1016/j.icarus.2004.07.031}.

\bibitem[{Greve et~al.(2004)Greve, Mahajan, Segschneider and
  Grieger}]{greve_etal_04a}
Greve, R., R.~A. Mahajan, J.~Segschneider and B.~Grieger. 2004.
\newblock Evolution of the north-polar cap of {Mars}: a modelling study.
\newblock \emph{{Planet. Space Sci.}}, \textbf{52}~(9), 775--787.
\newblock \doi{10.1016/j.pss.2004.03.007}.

\bibitem[{Grima et~al.(2009)Grima, Kofman, Mouginot, Phillips, H\'erique,
  Biccari, Seu and Cutigni}]{grima_etal_09}
Grima, C., W.~Kofman, J.~Mouginot, R.~J. Phillips, A.~H\'erique, D.~Biccari,
  R.~Seu and M.~Cutigni. 2009.
\newblock North polar deposits of {Mars}: {Extreme} purity of the water ice.
\newblock \emph{{Geophys. Res. Lett.}}, \textbf{36}, L03203.
\newblock \doi{10.1029/2008GL036326}.

\bibitem[{Haberle et~al.(2003)Haberle, Murphy and Schaeffer}]{haberle_etal_03}
Haberle, R.~M., J.~R. Murphy and J.~Schaeffer. 2003.
\newblock Orbital change experiments with a {Mars} general circulation model.
\newblock \emph{{Icarus}}, \textbf{161}~(1), 66--89.

\bibitem[{Head et~al.(2003)Head, Mustard, Kreslavsky, Milliken and
  Marchant}]{head_etal_03}
Head, J.~W., J.~F. Mustard, M.~A. Kreslavsky, R.~E. Milliken and D.~R.
  Marchant. 2003.
\newblock Recent ice ages on {Mars}.
\newblock \emph{{Nature}}, \textbf{426}~(6968), 797--802.
\newblock \doi{10.1038/nature02114}.

\bibitem[{Herkenhoff and Plaut(2000)}]{herkenhoff_plaut_00}
Herkenhoff, K.~E. and J.~J. Plaut. 2000.
\newblock Surface ages and resurfacing rates of the polar layered deposits on
  {Mars}.
\newblock \emph{{Icarus}}, \textbf{144}~(2), 243--253.

\bibitem[{Hvidberg(2003)}]{hvidberg_03}
Hvidberg, C.~S. 2003.
\newblock Relationship between topography and flow in the north polar cap on
  {Mars}.
\newblock \emph{{Ann. Glaciol.}}, \textbf{37}, 363--369.

\bibitem[{Ingersoll(1970)}]{ingersoll_70}
Ingersoll, A.~P. 1970.
\newblock Mars: Occurrence of liquid water.
\newblock \emph{{Science}}, \textbf{168}~(3934), 972--973.

\bibitem[{Jakosky(1985)}]{jakosky_85}
Jakosky, B.~M. 1985.
\newblock The seasonal cycle of water on {Mars}.
\newblock \emph{{Space Sci. Rev.}}, \textbf{41}, 131--200.

\bibitem[{James et~al.(1992)James, Kieffer and Paige}]{james_etal_92}
James, P.~B., H.~H. Kieffer and D.~A. Paige. 1992.
\newblock The seasonal cycle of carbon dioxide on {Mars}.
\newblock In: H.~H. Kieffer, B.~M. Jakosky, C.~W. Snyder and M.~S. Matthews
  (Eds.), \emph{Mars}, pp. 934--968. University of Arizona Press, Tucson, AZ,
  USA.

\bibitem[{Laskar et~al.(2004)Laskar, Correia, Gastineau, Joutel, Levrard and
  Robutel}]{laskar_etal_04}
Laskar, J., A.~C.~M. Correia, M.~Gastineau, F.~Joutel, B.~Levrard and
  P.~Robutel. 2004.
\newblock Long term evolution and chaotic diffusion of the insolation
  quantities of {Mars}.
\newblock \emph{{Icarus}}, \textbf{170}~(2), 343--364.
\newblock \doi{10.1016/j.icarus.2004.04.005}.

\bibitem[{Lewis et~al.(1999)Lewis, Collins, Read, Forget, Hourdin, Fournier,
  Hourdin, Talagrand and Huot}]{lewis_etal_99}
Lewis, S.~R., M.~Collins, P.~L. Read, F.~Forget, F.~Hourdin, R.~Fournier,
  C.~Hourdin, O.~Talagrand and J.-P. Huot. 1999.
\newblock A climate database for {Mars}.
\newblock \emph{{J. Geophys. Res.}}, \textbf{104}~(E10), 24177--24194.

\bibitem[{Michelangeli et~al.(1993)Michelangeli, Toon, Haberle and
  Pollack}]{michelangeli_etal_93}
Michelangeli, D.~V., O.~B. Toon, R.~M. Haberle and J.~B. Pollack. 1993.
\newblock Numerical simulations of the formation and evolution of water ice
  clouds in the {Martian} atmosphere.
\newblock \emph{{Icarus}}, \textbf{102}~(2), 261--285.

\bibitem[{Murray(1967)}]{murray_67}
Murray, F.~W. 1967.
\newblock On the computation of saturation vapor pressure.
\newblock \emph{{J. Appl. Meteorol.}}, \textbf{6}, 203--204.

\bibitem[{Plaut et~al.(2007)Plaut, Picardi, Safaeinili, Ivanov, Milkovich,
  Cicchetti, Kofman, Mouginot, Farrell, Phillips, Clifford, Frigeri, Orosei,
  Federico, Williams, Gurnett, Nielsen, Hagfors, Heggy, Stofan, Plettemeier,
  Watters, Leuschen and Edenhofer}]{plaut_etal_07}
Plaut, J.~J., G.~Picardi, A.~Safaeinili, A.~B. Ivanov, S.~M. Milkovich,
  A.~Cicchetti, W.~Kofman, J.~Mouginot, W.~M. Farrell, R.~J. Phillips, S.~M.
  Clifford, A.~Frigeri, R.~Orosei, C.~Federico, I.~P. Williams, D.~A. Gurnett,
  E.~Nielsen, T.~Hagfors, E.~Heggy, E.~R. Stofan, D.~Plettemeier, T.~R.
  Watters, C.~J. Leuschen and P.~Edenhofer. 2007.
\newblock Subsurface radar sounding of the south polar layered deposits of
  {Mars}.
\newblock \emph{{Science}}, \textbf{316}~(5821), 92--95.
\newblock \doi{10.1126/science.1139672}.

\bibitem[{Pollack et~al.(1990)Pollack, Haberle, Schaeffer and
  Lee}]{pollack_etal_90}
Pollack, J.~B., R.~M. Haberle, J.~Schaeffer and H.~Lee. 1990.
\newblock Simulation of the general circulation of the {Martian} atmosphere {1.
  Polar} processes.
\newblock \emph{{J. Geophys. Res.}}, \textbf{95}~(B2), 1447--1473.

\bibitem[{Read et~al.(1997)Read, Collins, Forget, Fournier, Hourdin, Lewis,
  Talagrand, Taylor and Thomas}]{read_etal_97}
Read, P.~L., M.~Collins, F.~Forget, R.~Fournier, F.~Hourdin, S.~R. Lewis,
  O.~Talagrand, F.~W. Taylor and N.~P.~J. Thomas. 1997.
\newblock A {GCM} climate database for {Mars:} for mission planning and for
  scientific studies.
\newblock \emph{{Adv. Space Res.}}, \textbf{19}, 1213--1222.

\bibitem[{Richardson and Wilson(2002)}]{richardson_wilson_02}
Richardson, M.~I. and R.~J. Wilson. 2002.
\newblock A topographically forced asymmetry in the martian circulation and
  climate.
\newblock \emph{{Nature}}, \textbf{416}~(6878), 298--301.
\newblock \doi{10.1038/416298a}.

\bibitem[{Sears and Moore(2005)}]{sears_moore_05}
Sears, D. W.~G. and S.~R. Moore. 2005.
\newblock On laboratory simulation and the evaporation rate of water on {Mars}.
\newblock \emph{{Geophys. Res. Lett.}}, \textbf{32}~(16), L16202.
\newblock \doi{10.1029/2005GL023443}.

\bibitem[{Segschneider et~al.(2005)Segschneider, Grieger, Keller, Lunkeit,
  Kirk, Fraedrich, Rodin and Greve}]{segschneider_etal_05}
Segschneider, J., B.~Grieger, H.~U. Keller, F.~Lunkeit, E.~Kirk, K.~Fraedrich,
  A.~Rodin and R.~Greve. 2005.
\newblock Response of the intermediate complexity {Mars Climate Simulator} to
  different obliquity angles.
\newblock \emph{{Planet. Space Sci.}}, \textbf{53}~(6), 659--670.
\newblock \doi{10.1016/j.pss.2004.10.003}.

\bibitem[{Smith et~al.(1999)Smith, Zuber, Solomon, Phillips, Head, Garvin,
  Banerdt, Muhleman, Pettengill, Neumann, Lemoine, Abshire, Aharonson, Brown,
  Hauck, Ivanov, McGovern, Zwally and Duxbury}]{smith_etal_99a}
Smith, D.~E., M.~T. Zuber, S.~C. Solomon, R.~J. Phillips, J.~W. Head, J.~B.
  Garvin, W.~B. Banerdt, D.~O. Muhleman, G.~H. Pettengill, G.~A. Neumann, F.~G.
  Lemoine, J.~B. Abshire, O.~Aharonson, C.~D. Brown, S.~A. Hauck, A.~B. Ivanov,
  P.~J. McGovern, H.~J. Zwally and T.~C. Duxbury. 1999.
\newblock The global topography of {Mars} and implications for surface
  evolution.
\newblock \emph{{Science}}, \textbf{284}~(5419), 1495--1503.

\bibitem[{Stenzel et~al.(2007)Stenzel, Grieger, Keller, Greve, Fraedrich, Kirk
  and Lunkeit}]{stenzel_etal_07}
Stenzel, O.~J., B.~Grieger, H.~U. Keller, R.~Greve, K.~Fraedrich, E.~Kirk and
  F.~Lunkeit. 2007.
\newblock Coupling {Planet Simulator Mars}, a general circulation model of the
  {Martian} atmosphere, to the ice sheet model {SICOPOLIS}.
\newblock \emph{{Planet. Space Sci.}}, \textbf{55}~(14), 2087--2096.
\newblock \doi{10.1016/j.pss.2007.09.001}.

\bibitem[{Takahashi et~al.(2003)Takahashi, Fujiwara, Fukunishi, Odaka, Hayashi
  and Wanabe}]{takahashi_etal_03}
Takahashi, Y.~O., H.~Fujiwara, H.~Fukunishi, M.~Odaka, Y.-Y. Hayashi and
  S.~Wanabe. 2003.
\newblock Topographically induced north-south asymmetry of the meridional
  circulation in the {Martian} atmosphere.
\newblock \emph{{J. Geophys. Res.}}, \textbf{108}~(E3), 5018.
\newblock \doi{10.1029/2001JE001638}.

\bibitem[{Tillman(2001)}]{tillman_01}
Tillman, J.~E. 2001.
\newblock Mars: Temperature overview.
\newblock Online publication.
\newblock \urlprefix\url{http://www-k12.atmos.washington.edu/k12/}, retrieved
  2009-07-15.

\bibitem[{Winebrenner et~al.(2008)Winebrenner, Koutnik, Waddington, Pathare,
  Murray, Byrne and Bamber}]{winebrenner_etal_08}
Winebrenner, D.~P., M.~R. Koutnik, E.~D. Waddington, A.~V. Pathare, B.~C.
  Murray, S.~Byrne and J.~L. Bamber. 2008.
\newblock Evidence for ice flow prior to trough formation in the martian north
  polar layered deposits.
\newblock \emph{{Icarus}}, \textbf{195}~(1), 90--105.
\newblock \doi{10.1016/j.icarus.2007.11.030}.

\bibitem[{Zuber et~al.(1998)Zuber, Smith, Solomon, Abshire, Afzal, Aharonson,
  Fishbaugh, Ford, Frey, Garvin, Head, Ivanov, Johnson, Muhleman, Neumann,
  Pettengill, Phillips, Sun, Zwally, Banerdt and Duxbury}]{zuber_etal_98}
Zuber, M.~T., D.~E. Smith, S.~C. Solomon, J.~B. Abshire, R.~S. Afzal,
  O.~Aharonson, K.~Fishbaugh, P.~G. Ford, H.~V. Frey, J.~B. Garvin, J.~W. Head,
  A.~B. Ivanov, C.~L. Johnson, D.~O. Muhleman, G.~A. Neumann, G.~H. Pettengill,
  R.~J. Phillips, X.~Sun, H.~J. Zwally, W.~B. Banerdt and T.~C. Duxbury. 1998.
\newblock Observations of the north polar region of {Mars} from the {Mars
  Orbiter Laser Altimeter}.
\newblock \emph{{Science}}, \textbf{282}~(5396), 2053--2060.

\end{thebibliography}

\end{small}

\end{document}